\documentclass[12pt]{article}

\usepackage{amssymb,fullpage,graphicx,url}

\newcommand{\R}{{\mathbb R}}
\newcommand{\N}{{\mathbb N}}
\newcommand{\C}{{\mathbb C}}
\newcommand{\Z}{{\mathbb Z}}
\newcommand{\cH}{{\cal H}}

\newcommand{\cA}{{\cal A}}

\newtheorem{Theorem}{Theorem}
\newtheorem{Lemma}{Lemma}
\newtheorem{Definition}{Definition}
\newtheorem{Example}{Example}

\newtheorem{Question}{Question}

\date{September 09, 2010}

\begin{document}

\title{Is  there a physically universal cellular automaton or Hamiltonian?}

\author{Dominik Janzing\\{ \small Max Planck Institute for Biological Cybernetics}\\
{\small Spemannstr.~38}\\
{\small 72076 T\"ubingen,  Germany}\\
{\small Email: \texttt{dominik.janzing@tuebingen.mpg.de}}}

\maketitle

\abstract{It is known that both quantum and classical cellular automata (CA) exist that are
computationally universal in the sense that they can simulate, after appropriate initialization,
any quantum or classical computation, respectively. Here we  introduce a different notion of universality: a  CA is called physically universal if every transformation on any finite region can be (approximately) implemented by the autonomous time evolution of the system after 
 the complement of the region has been initialized in an appropriate way.
We pose the question of whether physically universal CAs exist. 

Such  CAs would provide a model of the  world where the boundary between a physical system
and its controller can  be consistently shifted, in analogy to the Heisenberg cut for the 
quantum measurement problem.  
We propose to study the thermodynamic cost of computation and control
within such a model because 
implementing a cyclic process on a microsystem may require
a non-cyclic process for its controller, whereas implementing a cyclic process on system {\it and} controller
may require the implementation of a non-cyclic process on a ``meta''-controller, and so on. 
Physically universal CAs avoid this infinite hierarchy of controllers and  
the cost of implementing cycles on a subsystem can be described by mixing properties of the CA dynamics.

We define a physical prior on the CA configurations by  applying the dynamics  to
an initial state  where
half of the CA is in the maximum entropy state and half of it is in the all-zero state (thus reflecting the fact that life requires non-equilibrium states like the boundary between a hold and  a cold reservoir). 
As opposed to Solomonoff's prior, our prior  
does not only account for the
Kolmogorov complexity but also for the cost of isolating the system during the state preparation if 
the preparation process is not robust. 

The main goal of this article is to formally state several open problems and sketch their relevance
for the foundations of physics
rather than providing results.
}

\section{Towards a physical theory of control}

In  the abstract framework of both quantum theory and classical physics,
the following  concepts play a crucial role:
(1) states (2) dynamical evolution (3) measurements (4) system composition and (5) restriction  of the  state of a composed system to one of its components. 
In  quantum theory, 
states are given  by density operators (e.g. positive operators with trace one)  
on the system Hilbert space $\cH$, the dynamical evolution is described by 
a semi-group of completely positive trace-preserving maps, measurements are described by positive-operator-valued measures, and system composition is described  by tensor products of Hilbert spaces
\cite{Ja68,Da76,NC}.  Finally, partial  traces define system  restriction.

In classical physics, the states are probability distributions on a phase space,
the dynamics is given by a semi-group of stochastic maps, system composition is given by the cartesian product of the phase spaces, and state restriction is given  by marginalization of probability measures.

Having such a framework for the  physical world  raises the question to what  extent the formalism 
also contains states, dynamical evolutions, and measurements that do not correspond to any physically  possible situation or process. 
Restricting the attention to quantum theory, these questions thus read:
(1) Is every density operator on $\cH$ a {\it physically} possible state, (2) is every completely positive trace-preserving operation a process that can be implemented in nature,
(3) is there a measurement procedure for every POVM? 

First we describe in what sense modern quantum computing  (QC) research \cite{NC} has given an affirmative answer to all  these questions and in what sense it has not. To this end, we first rephrase some terminology
of QC. A quantum-bit (qubit) is a quantum system with Hilbert space $\C^2$, a quantum  register is a collection of $n$ qubits\footnote{It should be noted that the restriction to two-dimensional systems is only a matter of convention.}.  Researchers have described various physical systems having 
a  quantum degree of  freedom for which two states are universally controllable in  the following  sense:
Any unitary operation on $\C^2$ (``single qubit gate'') can be performed by appropriate operations on the system. Moreover,
they have described  how to implement controlled interactions between pairs  of qubits, thus implementing
a unitary on  the Hilbert space $\cH:=\C^2\otimes \C^2$ that is not a product of single qubit operations
(hence a proper ``two-qubit gate'').
It was then shown that sequences of one- and two-qubit gates are sufficient for implementing arbitrary unitary operations
up to   any  desired precision \cite{DiV}. Being able to prepare {\it one} pure state of the quantum register thus enables the preparation of {\it any} pure state. 
Moreover, measurements with respect to any measurement basis can be reduced 
to measurements with respect to a single reference basis by first transforming the state to the latter 
basis via a unitary transformation. 
Preparations of mixed states, implementation of general
completely positive trace-preserving maps and measurements for general POVMs can be obtained by
restriction of states to a subsystem. In this sense, questions (1)-(3) seem to  be answered  with ``yes''.
Then, the operational meaning of some multi-qubit states, dynamical evolutions, observables are only limited by the  fact that the  implementation time  could even exceed the life-time of the universe\footnote{For a complexity theory of states and observables see e.g. \cite{JB00c,Soklakov,PSPACE_QIC,Habil}}.
There are, however,  two other reasons why QC did not answer our questions in the sense intended here.

First, the quantum mechanical  degrees of freedom defining the qubits
in existing proposals for QC \cite{NC}
are only a small part
of the entire degrees of freedom  of physical particles  
(e.g. the nuclear spin of a particle or it may be two levels in an internal degree of freedom of a trapped ion).
So far, it has not been claimed that {\it all} the  degrees of freedom of such a particle would be controllable simultaneously. 

The second reason why we are not satisfied  with the answer given by QC is that
we would like to see a theoretical model of quantum control
that treats the controller as the same type of physical system as the system
to be controlled.
Within such a unifying model -- as proposed by the present article -- we are able to explore the conditions under which
one system acts as controller of the other, even though a physical interaction can
send information in {\it both} directions.\footnote{Note that unidirectionality
of causal influence not only occurs if the controller is significantly larger than the system
to be controlled. Instead, it is also a matter of the {\it state} of the controller. For such toy models of quantum control see e.g., \cite{JZAB,controlled_controller}; Refs.~\cite{heat_information,MitArmen} discuss thermodynamic  aspects of unidirectionality.}
Moreover, the question of how to control the  controller then   
shifts the problem of how to control the system 
to the problem of how to control the controller by a ``meta-controller'', leading to an infinite  
hierarchy of controllers.
 
 Remarkably, the same shift between system and its interface is generally accepted for
 the quantum measurement problem:  Once the quantum measurement process is described by an interaction between system  and
the measurement apparatus, the question occurs ``who measures  the measurement apparatus?'', which leads to the same chain of
measurement instruments as we have stated for the controller  problem  above.
For the measurement apparatus, it has been argued that the cut between system  and measurement instrument is arbitrary, the description must remain consistent if the boundary is shifted. Likewise, we argue that quantum control has a consistent description 
if one can show that the cut between system  and controller can be shifted.  In \cite{JZAB}, we have already described a toy model of quantum control with a fixed interaction between
controller and system, where operations on the system are implemented by implementing transformations on the
controller. In the present paper, we assume that we are only able to implement {\it state preparations} on the controller. We first state on an abstract level what we would consider a consistent model
of physical control,
 before it will be made precise within the
setting of cellular automata (CA):

\begin{Definition}[model of physical control, abstract version ]${}$\\ \label{PC}
Let $(\alpha_t)$ with $t\in \R$ or $t\in \Z$ be a group describing the dynamical
evolution on 
state space of the world $W$. 
Then every mathematically possible operation on the physical state space of some region $R$
can be implemented by initializing the complement $W\setminus R$ of the region to an appropriate state
and waiting until $\alpha_t$ implements the desired operation.
\end{Definition}

To motivate Definition~\ref{PC}, we first consider an arbitrary experimental setup that
is able to implement one particular control operation. The control operation may, for instance, be
to change the quantum state of a few ions in an ion trap in some desired way. To this end, some sophisticated
sequence of Laser pulses is applied to the system. Assume that the pulses are controlled 
by a computer program so that there  is no need  for the experimentalist to intervene once the program runs. 
We can then consider computer, Laser and the ions in the trap as a big physical system on which the global 
dynamics of the world acts. Obviously, the computer software controlling this process
is just a physical state of the computer. 
However, here we want to go further and also consider 
the presence or absence of the hardware of the experimental setup merely as different states of a larger  system (which implicitly refers to a field-theoretic point of view).
From such a perspective,  
 there is no distinction between hardware and software in the experimental setup 
and
the whole control operation on  the system to be controlled (the ions) is implemented by changing the physical state of the system's environment. 

Following, for instance,  \cite{Zuse,Fredkin,Cheung} we will consider
cellular automata (CA) as interesting models of the world  and therefore study 
our problem in the context of CAs. 
Our main focus (Section~\ref{Sec:classical}) will be on classical CAs  since the problem seems to be non-trivial even in the classical regime. Apart from describing possible  definitions of physical
universality (Subsection~\ref{Subsec:def}), we discuss some relations between physical universality 
to ergodic properties of CAs in Subsection~\ref{Subsec:ergodic}.  Subsections~\ref{Subsec:limits} and \ref{Subsec:thermo}
argues why physically universal CAs are helpful for  studying limits of control and 
thermodynamic laws from a new perspective.
Subsection~\ref{Subsec:Kolmo} proposes a prior distribution for physical states
based on physically universal CAs. To  this end, we consider  an initial state of the CA
where  half of the cells are set  to zero and the other half are in the maximum entropy state
(thus modelling a hot and a  cold  part of the  universe).
 Section~\ref{Sec:quantum}
briefly discusses physical universality for {\it quantum} CAs (Subsection~\ref{Sec:quantum}) and physically universal Hamiltonians as  their continuous analog (Subsection~\ref{Subsec:Hamil}), where controllability also
implies the ability to control the preparation of quantum  superpositions by a {\it classical} program. 
In the context of physically universal Hamiltonians,  the terms ``hot'' and ``cold'' part  of the universe
can be taken more literally because they really refer to Gibbs states. This makes the  physical interpretation of the
prior more obvious. 

The main  contribution of this  article is to raise the question of 
how to define the right  framework for a physical control theory  
that also treats the controller as an object internal to the theory. 
In posing this question, the paper sketches the possible impact of such a framework, but
it will not present any deep results on cellular automata.

\section{Physically universal classical cellular automata}

\label{Sec:classical}

\subsection{Possible options for defining physical universality}

\label{Subsec:def}

We first  introduce some terminology and  notation for classical cellular automata (CAs).
Let $L:=\Z^d$ for some $d\in \N$ be a $d$-dimensional lattice and
$A$ be an alphabet of states of a single cell (without loss of generality, let one of the symbols be ``$0$'').  
 The space of pure states of the CA is given by
\[
S:=A^L\,.
\]
The space of mixed states is given  by probability distributions  on  $S$.
The maximally mixed state   (maximum entropy state) is given by the uniform  distribution over $S$, i.e., the infinite  product  
of uniform  distributions over $A$.
For every state $s\in S$ and every subset $R\subset L$ the restriction of $s$ to $R$, denoted
by $s|_R$  
is defined by the substring  $s'\in A^R$ corresponding to $R$.
A {\it region}  will be a subset $R\in L$. Usually, our regions will be finite subsets unless
we state the opposite.

By slightly abusing notation, we set
\[
s|_x :=s|_{\{x\}}\,,
\]
for any $x\in L$.
A  configuration of a region $R$ is a string $c\in A^R$. It defines,  in a straightforward way, 
the cylinder set
\[
\{ s \in A^L \,  |\quad \, s|_R=c\}\,,
\]
which will also be denoted by $c$ whenever this causes no confusion.
The entropy of $R$ in the mixed state $\nu$ is given  by the Shannon entropy
of the restriction of $\nu$ to $R$, i.e.,
\[
S(\nu|_R):=- \sum_{c\in A^R} \nu(c)  \log  \nu (c)\,.   
\]

The time evolution $(\alpha_t)_{t\in \Z}$ of a CA is a group (by assuming the group property
we implicitly restrict the attention to reversible CAs)
 of translation covariant maps 
\[
\alpha_t: S \rightarrow S\,,
\]
that is local in the sense that $\alpha_{\pm 1}  (s)|_x$ only depend on the state of the cells
lying  in some  neighborhood of $x$.
Here we consider the Moore neighborhood of radius one, i.e., all cells $y$ with $\|y-x\|_\infty\leq 1$
\cite{Kari}. 

By slightly overloading notation, we also write $\alpha_t(c)|_R$ if $c\in A^{R'}$  is  the configuration
of any region $R'$ that contains all cells   relevant  for determining
the state of $R$ at time $t$ (which is, for instance, the case if $R'$ contains
the Moore neighborhood of $R$  with radius $t$).
If a configuration $c\in A^R$ is defined via $c:=(c_1,c_2)$ with $c_1\in A^{R_1}$ and
$c_2\in A^{R_2}$ for $R=R_1\cup R_2$, we write
$\alpha_t(c_1,c_2)|_R$ instead of $\alpha_t((c_1,c_2))|_R$. 

The following definition formalizes the weakest form among all notions of physical universality
that we define. It is the ability to change the state of a region $R$ by initializing the complement
of $R$  in an  appropriate way:

\begin{Definition}[conditional state preparation]${}$\\ \label{Cstate}
A CA is said to allow for conditional state preparation if
for every region $R\subset L$ and every pair $(c_i,c_f)$ of initial and final
configurations of $R$ there exists a configuration $e\in A^{L \setminus R}$ and a time $t\in \N_0$ 
such that
\[
\alpha_t (e,c_i)|_{R}=c_f \,.
\]
Less formally speaking, the dynamics prepares the final state $c_f\in A^R$ after the time $t$,  given that
 the environment started in the state $e$ and the region in the state $c_i$.
\end{Definition}

\noindent
Note that the state $e$ can be chosen differently for every initial state $c_i$. The following notion of state preparation is stronger since it demands the existence of a state $e$ that works for {\it  every} initial configuration  $c_i$:

\begin{Definition}[unconditional state preparation]${}$\\ \label{Ustate}
A CA is said to allow for unconditional state preparation if
for every finite subset $R\subset L$ of cells and configurations $c_f \in A^R$, 
there exists a 
configuration in $e\in A^{L\setminus R}$ and a time $t\in \N$ such that 
\[
\alpha_t (e,c_i)|_{R}=c_f \,,
\]
for every configuration $c_i \in A^R$.

Less formally speaking, the dynamics prepares the state $c_f$ in the region $R$ by initializing
the complement of $R$ to $e$, regardless of the initial state $c_i$ of $R$.
\end{Definition}

\noindent
It seems that  Definition~\ref{Cstate} already formalizes a sufficiently strong property because 
one could prepare the environment after having read out the initial state $c_i$ of the region $R$.
However, the entire process of readout and  conditioning the initialization  of the  complement of $R$
on the state $c_i$ should also be implemented by the  physical laws that govern  the dynamics of 
the world. Therefore, we consider the latter definition as the better notion of universal state preparation.
Nevertheless, the following example shows that Definition~\ref{Ustate} is a rather weak notion
of universality   since  it is already satisfied by a simple shift:

\begin{Example}[shift]${}$\\ \label{shift}
For $x\in L$ let $\alpha_1$ be given  by shifting the state by the vector $x$, i.e.,
\[
\alpha_1(s)|_i:=s|_{i-x} \quad \quad  \forall s\in A^L\,.
\]
For some finite region $R\subset\Z$, let $c_f\in A^R$ be an arbitrary configuration.
Then $c_f$ can be prepared as follows. Choose some $t_0$ such that
\[
(R+x t_0)  \cap R =\emptyset\,.
\]
Initialize the region $R':=R-x t_0$ to the translated copy of $c_f$. Then the region $R$ is obviously in the configuration $c_f$ at time $t_0$. 

If $d=1$ and $x=1$, the dynamics shifts the state of each site by one. Then the corresponding MDS is 
known as Bernoulli shift.
\end{Example}

However,
such a trivial model of dynamical evolution is unacceptable as a model for universal control.
One reason is that it lacks computation power. We could ask for models that are computationally universal and allow for universal state preparation in the sense of Definitions~\ref{Cstate}
 or \ref{Ustate}. Rather than postulating computational power a priori, we prefer
 demanding that the model allows for non-trivial operations other than state preparation.
 The following condition includes conditional state preparation and is obviously not satisfied for the shift dynamics:

\begin{Definition}[universal implementation of bijections]${}$\\ \label{bijections}
A CA is said to allow for universal implementation of bijections if for every finite region $R\subset L$ and every 
bijective map 
\[
\pi: A^R \rightarrow A^R
\] 
there is a configuration $e\in A^{L\setminus R}$ of the complement of $R$ and a time $t$ such that 
\[
\alpha_{t} (e,c)|_R=\pi(c) \quad \forall c \in A^R\,.
\]
\end{Definition}

Note  that the ability of implementing bijections implies the ability of implementing measurements
in the following sense: apart from a region $R$ whose state should be measured, define a  region $R_M$ 
which serves as a measurement aparatus. One can then implement a bijection $\pi$ on  $R\cup R_M$ that
chnages  the  state of $R_M$ depending on  the state of $R$.

One of the main goal of this paper  is to formulate  the following open problem:

\begin{Question}[existence of physically universal CA]${}$\\  \label{Ex}
Is there a classical CA that is physically universal in the sense of
Definition~\ref{bijections}?
\end{Question}

It  is easy to see that non-bijective maps $\pi$ can be implemented by restricting bijections
to smaller regions.  For this  reason, the bijectivity assumption in Definition~\ref{bijections} is irrelevant  and it is a matter of taste whether  one wants  to keep it in the definition.

In case the answer to this question is negative, one should try to find a  weaker sense of universal
controllability. An affirmative answer, on the other hand, raises further 
questions since physically universal CAs are good candidates for studying thermodynamic cost
of computation and (quantum) control from a new perspective. Some ideas on that will be presented
in Subsection~\ref{Subsec:thermo}.

We will not formulate any conjecture  regarding the solution of Question~\ref{Ex}, but 
Subsection~\ref{Subsec:limits} will show that the controllability of the controller of a  system imposes
limitations  on the controllability of the system itself.

\subsection{Some relations between physical  universality and ergodic  properties}

\label{Subsec:ergodic}

We want to discuss relations between physical universality and ergodicity of
dynamical systems.
To this end, we introduce the following terminology \cite{isem}:

\begin{Definition}[measure-preserving dynamical systems (MDS)]${}$\\
Let $(\Omega,\Sigma,\mu)$ be a measure space where $\Omega$ is a set, $\Sigma$ the $\sigma$-algebra
of measurable subsets of $\Omega$ and $\mu$ a measure with $\mu(\Omega)<\infty$.
Let $\phi: \Omega \rightarrow \Omega$ be a measurable map with $\mu(\phi^{-1}(B))=\mu(B)$
for every measurable  set $B$.
Then $(\Omega,\Sigma,\mu,\phi)$ is called a measure-preserving dynamical system (MDS).
\end{Definition}

\noindent
Then we have:

\begin{Lemma}[CA is an MDS]${}$\\
Every reversible CA as defined above is a measure-preserving dynamical system where
$\Omega:=S$, $\Sigma$ is generated by the set of cylinder sets, $\mu$ is the product of 
uniform probability distributions on $A$ and $\phi:=\alpha_1$.
\end{Lemma}

\vspace{0.3cm}
\noindent
Proof: $\mu( \alpha_1^{-1} (B)) =\mu (B)$ can easily be checked for every cylinder set $B$.
Since the latter ones generate the entire sigma algebra of measurable sets, conservation of measure
follows.$\Box$ 

\vspace{0.3cm}
\noindent
The following terminology will be useful \cite{Halmos,isem}:

\begin{Definition}[ergodicity]${}$\\
An MDS is called ergodic if $\phi^{-1} (B)\cong B$ implies 
$B\cong \Omega$ or $B\cong \emptyset$ for all $B\in \Sigma$, where $\cong$ denotes equality up to sets of measure zero. Equivalently, $\phi^{-1}(B)\cong B$ can also be replaced with
$\phi^{-1}(B)\subseteq B$ or $\phi^{-1}(B)  \supseteq B$ (up to sets of measure zero).
\end{Definition}

\noindent
We will also need another equivalent formulations of ergodicity \cite{Halmos}:

\begin{Lemma}[different characterization of ergodicity]${}$\\
An MDS is ergodic if and only if for every $B,D \in \Sigma$ 
there is an $t\in \N$ such that
\[
\phi^{-t}(B) \cap D \not\cong \emptyset\,.
\]
\end{Lemma}

\noindent
Then we have:

\begin{Theorem}[state preparation in ergodic CAs]${}$\\  
If a CA is an ergodic MDS, it allows for conditional state preparation
in the sense of Definition~\ref{Cstate}.
\end{Theorem}

\vspace{0.3cm}
\noindent
Proof: Let  $B_i\subset S$ and $B_f\subset S$ 
be the cylinder sets corresponding to the initial and the final configuration
$c_i$ and $c_f$ of $R$, respectively. 
Then there is a $t$ such that 
\[
\alpha^{-1}_t(B_f) \cap B_i \not\cong \emptyset\,.
\]
Choose $c\in \alpha^{-1}_t(B_f) \cap B_i$.
Since $c$ is an element of $B_i$,  it is of the form $c=(e,c_i)$.
On the other hand, $\alpha_t(e,c_i)|_R=c_f$
because $c\in \alpha_t^{-1}(B_f)$.
$\Box$

\vspace{0.3cm}
\noindent
Ergodicity of CAs has already been studied in the literature\footnote{Note that \cite{Richter}
studies ergodic quantum CAs, but not in the sense of MDS. Instead, ergodicity is meant in the sense
of a topological dynamics having a unique invariant state.}
 \cite{Shirvani,Willson}, but
the fact that the Bernoulli shift (Example~\ref{shift}) is ergodic \cite{isem} shows that 
even ergodicity does not imply physical universality in the sense  of  Definition~\ref{bijections}.

\subsection{Limits of controllability}

\label{Subsec:limits}

Being able to prepare a certain state, one may also wish to keep it at least for 
some time.
In the context of quantum information processing, for instance,  it  is considered as an important problem
to prevent a quantum state from decaying too quickly (where decay can be understood in the sense
of both decoherence or relaxation). To ensure this,  one tries to isolate the system as much as possible
from influences  of the environment. On the other hand, implementing control  operations
requires interactions with  the environment. 
We expect that this conflict between protecting the state by isolating the system and nevertheless still being 
able to access it, can be nicely explored in the setting of physically universal CAs. Then, 
isolating the system only means to prepare the environment into a state that effectively turns 
off the interaction. The question of whether this conflict implies serious restrictions
to physical universality will mainly be unanswered, but we 
mention
some small observations that may suggest a future direction for research.
The following statement, for instance, is almost obvious, but we phrase it as a theorem because
it shows that too strong controllability assumptions are self-contradictory:

\begin{Theorem}[some configurations are unstable]${}$\\
Let $R$ be a region that includes at least the Moore neighborhood of one cell  $x$.
Let $\alpha_t$ be physically universal in the sense of Definition~\ref{Cstate},
then there is a configuration $c\in A^R$ such that 
\[
\alpha_1(e,c) |_R \neq c \quad \forall e \in A^{L\setminus R}\,.
\]
\end{Theorem}

\noindent
Proof: If the dynamics of  the CA is non-trivial  (which is certainly the case
for physically universal CAs) there must be  a configuration $c\in A^R$ such that
\[
\alpha_1(c)|_x  \neq c|_x\,.
\]
Hence,
\[
\alpha_1(e,c)|_x \neq c_x
\]
for all $e\in A^{L\setminus R}$. 
$\Box$  

\vspace{0.3cm}
\noindent
The following result is only slightly less straightforward, but it already illustrates
how controllability of the controller of a region $R$ restricts 
the controllability of $R$:

\begin{Theorem}[no configuration lasts forever]${}$\\ \label{Th:not_forever}
Given a CA that is physically universal in the sense of Definition~\ref{bijections}, then
it is impossible that there exists  
initial and final configurations $c_i,c_f\in A^R$,  a finite ``program'' region $R_{p}$ with initialization $c_{p}$, and a time $t_0\in \N$ such that 
\begin{equation}\label{prepared_forever}
\alpha_t(c_i,c_{p}) |_R =c_f\quad \forall t \geq t_0\,.
\end{equation}
\end{Theorem}

\noindent
Proof: Assume that (\ref{prepared_forever}) is satisfied. 
Set $R':=R\cup R_p$ and choose a vector  $x\in L$  such that 
$(R'+x) \cap R' =\emptyset$ and that  $\|x\|_\infty \geq t_0$. Let $\beta$  be the transformation on $R' \cup (R'+x)$ that swaps the state between
$R'$ and $R'+x$. By physical universality in  the sense of Definition~\ref{bijections}, there is
a configuration of the complement of $R' \cup (R'+x)$ such that $\alpha_{t_1}$ implements
$\beta$ for some $t_1$. After the implementation of $\beta$, the region  $R$ is only 
in the state $c_f$ if the initial state of the region $R+x$ has been the shifted copy of
$c_f$. Hence, $t_1$ must be smaller than $t_0$ since (\ref{prepared_forever}) states that 
the state of $R$ is $c_f$  {\it regardless} of the state of $R'+x$ (note that $R'+x$  is part of  the
complement of $R_p$ by assumption and its state is thus irrelevant for (\ref{prepared_forever})).
On  the other hand, the implementation of the swap $\beta$ requires at least the time $t_0$
since the  information can propagate one cell per time step only, which leads to a contradiction.
$\Box$

\vspace{0.3cm}
Theorem~\ref{Th:not_forever} shows that initializing a finite region can never prepare a state that
lasts forever.  If possible at all,  it requires an {\it infinite} region. 
To show more powerful results about control tasks that are self-contradictory 
has to be left to the future (in this context it may also be worth mentioning 
Ref.~\cite{Wolpert} which describes some
impossibility results for 
{\it inference} tasks instead of control tasks within a  computation model of the world and relate them to the Halting  problem).

\subsection{Space and energy requirements of computations and control operations}
\label{Subsec:thermo}

In this section we want to mention some potential implications for the 
resource requirements of computation processes, given that physically universal CAs define
a reasonable model of the world. 
Even though we have proved only  a few results on this, 
the following  high-level arguments 
motivate 
why physically universal CAs shed a different light
on thermodynamics.

\begin{enumerate}

\item {\it The thermodynamic cost of isolating systems:} the difficulty of isolating physical objects from its environment is one of the main obstacles in
controlling microphysics. In usual quantum control, this appears more or less  as a  practical problem
and the question is  how to turn off the disturbing interactions. Physical universality,  however,
implies that the system is never isolated and that only appropriate states of the environment
ensure that the system behaves for some time  period as if it would be isolated. The fact  
that,  in turn,  also  the environment of the system is permanently coupled to its environment 
(by physical  universality)
implies
that this ``isolating state'' is perturbed after a while.  Preparing the environment into a  state
that effectively isolates the system for a long time, probably requires a lot  of thermodynamic resources.
To discuss these costs, one probably needs a model where all  interactions are permanently present
and cannot be turned on and off by the experimentalist. Within  the framework of 
physically  universal CAs  it  is not only possible to
address the requirements of extracting heat from a system
\cite{JWZGB}  but also of preventing the heat from reentering the system.

\item {\it Thermodynamic reversibility:} 
It is commonly assumed that the implementation of a bijective transformation of the  states
of a  microscopic system is  thermodynamically reversible.
The fact that the experimental setup controlling the implementation  generates a lot of  heat
is usually considered as a  problem of  current technology rather than being a fundamental law of physics.
Physically universal  CAs provide a model that makes  it possible to explore  how the controller 
(i.e., the region $R_p$ around the region $R$ to be controlled)
changes its  state during this implementation. 
From the point of view of traditional thermodynamics, this state transition
is again reversible if it is a  bijection of the state space of a  microsystem.
However, inverting this bijection will then change the state  of the environment
around $R_p$. Then, the question of
thermodynamic reversibility leads, again, to our infinite sequence of meta-controllers.
We  will  not present any solution to this deep problem. We only 
emphasize
 that the existence of thermodynamic reversible processes is challenged by the ideas  above.

\item {\it Space and energy requirements of computation:}
In complexity theory, the space requirements of a computation is defined as
the size of the memory band of a Turing machine that is written on during the
computation process. The complexity class PSPACE, for instance, is defined as the class of problems
whose space requirements increase only polynomial in the size of the input string \cite{Papa}.
It is known \cite{Berlekamp} that appropriate CAs can simulate a universal Turing machine efficiently with respect to both space and time resources. 

In our context, we want to redefine the space requirements of a computation in a way that
is motivated by ideas from thermodynamics: we do not only count those cells of the CA that are {\it actively} involved in the computation in the sense that their state changes during the process. Instead, we count all cells whose state matters. In the simplest case, it may be necessary to set a large set of cells to some fixed symbol (e.g. to zero) to avoid that these cells disturb the computation by influencing
the cells involved in the computation. From a purely computer scientific point of  view,
it is natural to study the resources of computation within a setting where all the sites are set to zero
except for those involved in the computation. In our physical model, however, this would correspond
to cooling all cells  down to zero temperature, which requires infinite thermodynamic resources.
 We assume that we can only  
extract the entropy of a {\it finite} region and use this free memory  space for the computation. 
In a physically universal CA, we then get  the problem that 
this region can never {\it remain} free of entropy
because the interaction that guarantees universality necessarily transfers entropy into
the free memory space. 

\end{enumerate}

The discussion below tries to support the vague statements above by formal arguments. 
We will 
not always distinguish between computation processes and
other control processes.\footnote{On the elementary level of nature,  thermodynamic 
and computation processes are closely  related, anyway \cite{HeatEngines}}.
The following theorem is actually a simple observation, but 
we  phrase it as  a  theorem because it confirms the last sentence of item 3 above:

\begin{Theorem}[lower bound on entropy influx]${}$\\
Let 
$R$ be an arbitrary region and $\nu$ be a probability  distribution on $S$ whose restriction to 
$L\setminus R$ is the uniform distribution.
Let the
CA be universal in the sense of Definition~\ref{bijections} and $x$ be some vector 
such that $R\cap  (R+x)=\emptyset$. 
If
$R_p$ denotes a region such that for some  $c_p  \in A^{R_p}$  
the state $c$ of $R$ is transferred to  $R+x$, i.e.,
\[
\alpha_t(c_p,c)|_{R+x} =c  \quad \forall c\in A^R\,,
\]
for some appropriate $t$, then 
the entropy of $R$ after  the time $t$ is  at least
\[
S((\nu \circ \alpha_t) |_R)\geq \frac{|R|}{|A|^{|R_p|}} \log |A|\,.
\]
\end{Theorem}

\noindent
Proof: For $\nu_t:=\nu  \circ \alpha_t$ 
we consider the conditional  distribution given 
$\alpha_t^{-1}(c_p)$. Its  restriction to $R$ is the uniform distribution because
the initial state $c_p$ triggers  the implementation of the swap between
$R+x$ and $R$. The entropy of the uniform  distribution   on $R$
reads $|R| \log |A|$. Since $R_p$ is initially also in 
the maximum entropy mixture,
the probability for being in the state $c_p$ is $|A|^{-|R_p|}$.
Weighting the entropy $|R| \log |A|$ with this factor yields the desired bound.
$\Box$

\vspace{0.3cm}
\noindent
The theorem shows a trade-off between being able to implement bijections 
and being able to isolate a region: if $\beta$ can be {\it easily} implemented  on  $R\cup (R+x)$ 
(i.e., by initializing a {\it small}  region $R_p$)  then 
$R\cup (R+x)$ is badly isolated because we get large entropy influx.
Note that no such statement holds for {\it computationally} universal CAs
since they could have a ``death state'' that 
remains forever and
turns off all interactions with the surrounding cells. 
A boundary with dead cells could then prevent the memory space 
from getting entropy from its environment.
In a physically universal CA, the environment is always able to ``revitalize'' the ``dead cells''.
 It is possible that in physically universal CAs, the region that needs to be initialized 
to enable a computation process
grows proportionally with the computation time.  Loosely speaking, the size of the
region that needs to be initialized is related to the amount of free energy
that must be available in order to run the computation properly.
This is because Landauer's principle  \cite{Landauer:61,BennettThermoReview,JWZGB}  states that it requires the energy
$E=kT \ln 2$ to initialize one bit. From a more accurate point of view, however,
we have to account for the fact that the region that we must initialize not necessarily needs 
to be prepared to {\it one} specific configuration. Instead, it  could be that there is a whole
set  of configurations that ensure that the desired computation process works properly.
This corresponds to a smaller amount of free energy.
The following definition formalizes the free energy content of configurations:

\begin{Definition}[free energy of a set of configurations]${}$\\ \label{free}
Let $B\subset A^L$ be a set of configurations  and $\mu$ be the uniform distribution
on $S$ (which is defined via the product of uniform distributions on each $A$).
Then
\[
F(B)= -\log_2 \mu(B) 
\]
is the free energy required to ensure that the world is in a state $s\in B$.
\end{Definition}

\noindent
The definition is motivated by the following interpretation of probability distributions.
The mixed state $\mu$, which is the uniform distribution over all configuration, is thought to be
the thermodynamic equilibrium of the world, i.e., the analog of the Gibbs state. We define its free energy to be zero.
In physics, the free energy of a mixed state is, up to the factor $kT$, given by
its relative entropy distance from thermal equilibrium \cite{OhyaPetz}.
Here, mixed states are probability distributions on $A^L$ and the free energy is thus
(up to constants that we ignore for sake of convenience)
given by the relative entropy distance from $\mu$, i.e.,
\[
F(\tilde{\mu}):= D(\tilde{\mu}||\mu)\,.
\]
If $\tilde{\mu}$ is any distribution with support $B$, the relative entropy distance to $\mu$ 
is minimal if $\tilde{\mu}$ is the uniform distribution on $B$. One checks easily that 
\[
D(\tilde{\mu}||\mu)=-\log_2 \mu(B)\,.
\]

Within  this setting, we can easily define the free energy needed for a preparation process:

\begin{Definition}[free energy required for a preparation process]${}$\\
Assume a region is in the state $c_i\in A^R$ and we want it to be in the state $c_f$ at time 
$t$.  Interpreting $c_i$ and $c_f$ as cylinder sets,  the state of the lattice $s\in A^{L\setminus R}$ 
must be chosen such that 
\[
s \in c_i \cap \alpha_t^{-1}(c_f)\,,
\]
where the right hand side interprets $c_i$ and $c_f$ as sets (as defined previously).

Then,
\[
F(c_i \mapsto c_f):=-\log \mu(c_i \cap \alpha_t^{-1}(c_f))
\]
is the  
free energy needed to implement the preparation process
$c_i \mapsto c_f$ after the  time $t$.
\end{Definition}

Note that this definition includes the free energy content of
$c_i$ which is given by $ |R|  \log_2 |A|$, since $c_i$ is   one configuration in a set  of
$|A|^{|R|}$ possible ones.

We also define the free energy required for a computation process:

\begin{Definition}[energy requirements for computation]${}$\\
Assume that the physical universal CA is only able to perform a desired computation process $C$ if 
the state $s$ of the world lies in the set $B \subset A^L$.
Then
\[
F(B):=-\log \mu(B)
\]
is the free energy required for $C$.
\end{Definition}

We will not elaborate on this any further, but consider the thermodynamic costs of implementing
{\it sequences of state transitions}  on some region
$R$ since this task is easier to address than computation tasks. 
Consider the following sequence of state transitions  
\[
c_0  \stackrel{t_1}{\mapsto} c_1 \stackrel{t_2}{\mapsto}  c_1 \cdots \stackrel{t_n}{\mapsto} c_n\,, 
\]
and define the corresponding free energy resource requirements by
\[
-\log \mu\left(c_0 \cap \alpha_{t_1}^{-1}(c_1) \cap \alpha_{t_1+t_2}^{-1} (c_2) \cap \cdots \cap \alpha^{-1}_{t_1+\cdots +t_n}(c_n)\right)\,.
\]
An interesting special instance is to implement $k$ cycles 
\begin{equation}\label{kcycles}
\underbrace{c_1  \mapsto c_2  \mapsto \cdots\mapsto c_n}_{\hbox{1th cycle}}\mapsto \underbrace{c_1 \mapsto c_2 \cdots c_n}_{\hbox{2nd cycle}} \cdots \,,  
\end{equation}
where the transition from $c_i$ and $c_{i+1}$ and  from $c_n$ to $c_1$ is implemented by 
one time step of the CA.  
We  do not know whether physically universal CAs also allow for the implementation of 
arbitrarily many cycles 
of this form, but given that they do, we have the following  statement for ergodic CAs:

\begin{Theorem}[cost of implementing repeated cycle processes]${}$\\
Let $c_1,\dots,c_n$ be configurations of a region $R$ such that
\begin{equation}\label{union}
\bigcup_{j=1}^n c_j \neq A^R\,.
\end{equation}
Then,
in an ergodic CA, the cost  of implementing $k$ cycles of the form (\ref{kcycles})
converges to infinity for $k\to\infty$.
\end{Theorem}

\noindent
Proof: define $B:=\bigcup_{j=1}^n c_j$ and 
\[
D:=\bigcap_{j=1}^k \alpha^{-1}_k (B)\,.   
\]
Clearly, 
\begin{equation}\label{inv}
\alpha_1(D)\subset D\,.
\end{equation} 
Due to eq.~(\ref{union}), we have $\mu(D)\neq 1$. Since $\alpha_1$ is ergodic,
all sets satisfying  the invariance condition (\ref{inv}) have measure zero or one, hence 
$\mu(D)=0$. Due to
\[
\lim_{k\to \infty} \mu\left(\bigcap_{j=1}^k \alpha_j (B)\right) =  \mu(D)=0\,,
\]  
the statement follows.
$\Box$.

A weaker task than implementing a cycle 
is to periodically 
restore the same configuration $c$ again and again after $\tau$ time steps,
without specifying what happens between the $\tau$ steps:
\[
c \stackrel{\tau}{\mapsto} c \stackrel{\tau}{\mapsto} c \stackrel{\tau}{\mapsto} \cdots \,.
\]
According to Definition~\ref{free}, 
the free energy requirements are given by
\begin{equation}\label{Cycles}
-\log \mu \Big( \bigcap_{j=0}^n \alpha_{j\tau}^{-1}(c) \Big) \,.
\end{equation}
To derive statements on the resources needed, we first
recall the following mixing property (see \cite{Halmos}, page 38), which is 
known to imply ergodicity \cite{Halmos,isem}:

\begin{Definition}[weakly mixing MDS]${}$\\
An MDS is called weakly mixing if 
\[
\lim_{n\to\infty} \frac{1}{n} \sum_{j=0}^{n-1}\mu\left( \phi^{-j} (B) \cap D\right) = \mu(B) \mu(D)\,,
\]
for all measurable sets $B,D$.
\end{Definition}

The following result of ergodic theory (Corollary 14.15 in \cite{isem}) will be helpful:
\begin{Lemma}[mixing of all orders]${}$\\
Every weakly mixing MDS is weakly mixing of all orders in the sense that
\begin{equation}\label{weakM}
\lim_{n\to \infty} \frac{1}{n} \sum_{n=0}^{n-1} \mu \left( B_0 \cap \phi^{-n}(B_1) \cap \phi^{-2n}(B_2) 
\cap \cdots \cap \phi^{-(k-1)n}(B_{k-1})\right)= \mu (B_0) \cdots \mu (B_{k-1}) \,,
\end{equation}
for all $k\in \N$ and every $B_0,\dots,B_{k-1}\in \Sigma$.
\end{Lemma}
We apply this result to our setting and obtain: 
\begin{Theorem}[cost of restoring states in weakly mixing CAs]${}$\\ \label{cycle_cost}
Let the CA be weakly mixing and assume that there  is a configuration $c\in A^R$  
for  which 
it is possible to implement
the following $k$-fold recurrence 
\[
\underbrace{c \stackrel{\tau}{\mapsto} c \stackrel{\tau}{\mapsto} c \mapsto \cdots \stackrel{\tau}{\mapsto} c}_k\,,
\]
for all $\tau\geq \tau_0$ for some $\tau_0 \in \N$.
Let $F_k(\tau)$ be the free energy required for implementing this process.
Define the average free energy requirements over all $\tau\geq \tau_0$ by
\begin{equation}\label{limit}
\bar{F}_k:=\liminf_{\tau_1\to \infty} \frac{1}{\tau_1-\tau_0+1} \sum_{\tau=\tau_0}^{\tau_1} F_k(\tau)\,.
\end{equation}
Then it satisfies  the lower  bound 
\[
\bar{F}_k \geq - k \log_2 \mu (c)\,. 
\]
\end{Theorem}

\noindent
Proof: According to  Definition~\ref{free}, $F_k(\tau)$ reads
\[
F_k(\tau):= -\log_2  \mu \left(\bigcap_{j=0}^{k-1} \alpha^{-1}_{j\tau} (c) \right)\,.
\]
Hence,
\[
\bar{F}_{k}:=- \liminf_{\tau_1\to \infty} \frac{1}{\tau_1-\tau_0+1}\sum^{\tau_1}_{\tau= \tau_0}  \log_2  \mu \left(\bigcap_{j=0}^{k-1} \alpha^{-1}_{j\tau} (c) \right)\,.
\]
The convexity of  the  logarithm implies
\[
\bar{F}_{k} \geq \log_2 \lim_{\tau_1\to\infty} 
\frac{1}{\tau_1-\tau_0+1}\sum^{\tau_1}_{\tau= \tau_0}  \mu \left(\bigcap_{j=0}^{k-1} \alpha^{-1}_{j\tau} (c) \right) = -\log_2 \mu(c)^k= -k \log_2 \mu(c)\,,
\]
where the second last equality uses eq.~(\ref{weakM}).
$\Box$

\vspace{0.3cm}
Theorem~\ref{cycle_cost} states that the cost of  repeatedly restoring the same state
$k$ times (after $\tau$ time steps) grows linearly in  $k$ when averaged over all $\tau$. 
The physical relevance of this statement is speculative for two reasons. 
First, we do not know whether the 
appropriate mixing properties follow from physical universality. Second, it  is unclear whether 
the  assumption that
the sequence  of state transitions can be  implemented for all  $\tau \geq \tau_0$ is reasonable.
We will therefore formulate another open problem:

\begin{Question}[thermodynamic cost of cycles]${}$\\
Given any desired configuration $c$, how does the free energy~(\ref{Cycles})
of restoring it again and again grow with the number  $k$ of cycles?
\end{Question}

In case 
the energy
grows at least linearly in $k$ for physically realistic models, this would 
suggest
that implementing cycles on microscopic systems involves 
an
 experimental setup whose energy content grows linearly in the number of cycles. 
On the one hand, the energy content does not seem to be {\it used up}, since it just needs  to be  {\it available}. On the 
other hand, this amount of energy cannot be used to implement the next cycles because, if reusing 
the energy was possible, the amount of energy that needs to be present would not grow linearly in $k$. 
Note that  the above  ergodic theory based framework avoids exploring  the thermodynamic cost of an infinite  sequence
of controllers and meta-controllers
as sketched in item 2 at the beginning of  this subsection
because the universal CA describes the whole  hierarchy of controllers simultaneously.

\subsection{Towards a physical analog of Kolmogorov complexity and Solomonoff's prior}

\label{Subsec:Kolmo}

Several authors have already pointed out the physical relevance  of algorithmic information (``Kolmogorov complexity''), e.g.,  \cite{Vitanyi08,Zu90}. For any binary string  $\{0,1\}^*$, the algorithmic information $K(s)$ is defined by
the length of the shortest program on a universal prefix Turing machine that outputs $s$ and  halts then
\cite{KolmoOr,Solomonoff2,Chaitin}.

The thermodynamic relevance of Kolmogorov complexity has, for instance, 
been emphasized in \cite{ZurekKol,Vitanyi08,MoraKraus}, its importance 
for statistical inference 
has already been described by Solmonoff \cite{Solomonoff2}, and also 
the foundation of modern machine learning 
methodology  often refer to Kolmogorov complexity, e.g. \cite{Hutter,Gruenwald}. 
Recently,  
\cite{LemeireD,LemeireJMLR2010,Algorithmic} postulated {\it causal} inference rules that also use algorithmic information. A crucial concept for algorithmic information based inference
is Solomonoff's prior:
\begin{Definition}[Solomonoff's prior]${}$\\
Given a  universal Turing  machine $T$ with prefix coding. Then,  for any binary string $s\in \{0,1\}^*$, one
defines $m(s)$ by the  probability that $T$ produces the output  $s$ and stops after every  bit of the infinite input tape
has been randomly set to  $0$ or $1$ with probability $1/2$ each.
\end{Definition}

Note that these random programs do not contain any {\it additional} symbol that indicates the end
of the program code. Since the Turing machine uses prefix coding, no valid program is the prefix of another one. For this reason, the uniform distribution over {\it all} binary words 
(defined by the infinite product  of uniform distributions on  $\{0,1\}$)
automatically defines 
a distribution on the set of valid programs.

Even though Solmonoff's prior has shown to be a powerful concept for the foundation of inference,
the following modifications may be appropriate for 
a prior on the states of the physical world:

\begin{enumerate}

\item {\it Symmetries:}  
What prior  probability should, for instance,  be assigned  to the event that a next lightening hits the 
earth at  a longitude of $0^o$ (up to an error of $\epsilon$)? 
There is no reason why it  should be larger than the probability of hitting  the earth at $24.35219^o$, because
nature does not care about whether the  numerical value of the  location can be computed  by a short  program. 
The physical laws  that govern lightening fulfill some symmetries  that should be respected by
our prior.  To construct  a prior that accounts for these symmetries and still captures the
aspect of description  length, we propose to use a computation model that 
is inherently symmetric with respect to some transformations.

\item {\it Complexity of isolating systems:} 
According to Solomonoff's prior, any state  having a short program as description is likely to occur  in nature, no matter whether the running time is large or not and no matter how robust the   output is
with respect to perturbing the state of the Turing machine during the computation. 
Physical prior probability should also account  for the {\it robustness} of 
the computation process since no system is perfectly isolated from its environment.
Physically universal CAs are good models to take  this into account  because
the coupling between system and its environment is always present by definition.

\end{enumerate}

We now define
a prior via a  physically universal CA. A naive analog of randomizing the input of the Turing machine would be to
initialize the CA to the  uniform distribution over all pure states  
and then applying the dynamics $\alpha_t$, yields a trivial prior
for every $t$ since our bijective dynamics preserves the uniform distribution. We want to define a prior that gives higher probability to {\it simple} patterns like $0^R$ (all cells in $R$ are in the state $0$).  
It will therefore be based on the following initial state:
\begin{Definition}[initial state of the universe]${}$\\ \label{initial}
Let $L=L_+ \cup L_-$ be a partition of the lattice into two infinite subsets
($L_+$ could,  for instance, be all cells with  $x_1>0$).
Define a probability measure $Q$ by setting all sites in $L_-$ to zero
and choosing the uniform distribution on $A^{L_+}$ (i.e. 
for  every site in $L_+$, a symbol is chosen independently with probability $1/|A|$ each).
\end{Definition}
We consider $L_+$ and $L_-$ as hot and cold parts of the world, respectively.
Then, interesting structure can only start growing at the boundary between hot and cold regions.
This accounts for the fact that life requires thermal   non-equilibrium, which is most naturally provided by temperature gradients.

Such a state ensures the availability of an infinite amount of free memory space.
-- A similar convention would also be required for
Solomonoff's prior if it was defined with respect to a {\it reversible} Turing machine \cite{Bennett:89}.
Then one would
also need to provide free memory space for free in order to ensure that
the string $0^k$ obtains a higher prior probability than a typical $k$-bit string.
We now define:
\begin{Definition}[physical prior]${}$\\ \label{Prior}
For every time $t\in  \N$, let 
$P_t$ be the probability distribution on $S$ that is obtained by
applying $\alpha_t$ to the initial mixed state $Q$, as given by
Definition~\ref{initial}. 
\end{Definition}

Let us discuss some properties of $P_t$.
As opposed to Solmonoff's prior, it depends on $t$. This is because
the Turing machine stops for appropriate inputs whereas the dynamics of our CA does not.
It is not clear whether one  should consider this as a feature
rather than as a drawback of our definition -- one may argue that in the early stage of the universe
other states were more likely than today and others were less likely.
Note, however, that 
\[
P(c) \propto \sum_{t\in \N_0} P_t(c) 2^{-K(t)}
\]
would be an option to define a time-independent prior. 
To elaborate on this  goes beyond the scope of  this  paper,  but the
additional term $K(t)$ will also appear in our definition of physical complexity
below. 

A second feature of $P_t$ is 
that the prior probability of a configuration (and also the physical complexity that we define below) depends on its location on the lattice: creating a cold region in the middle of the 
hot region involves much more sophisticated initialization than creating it close to the boundary to the cold region. In the former case, the entropy of the hot region needs to be transported over a long way to the cold region. 

\begin{figure}
\centerline{
\includegraphics[width=0.32\textwidth]{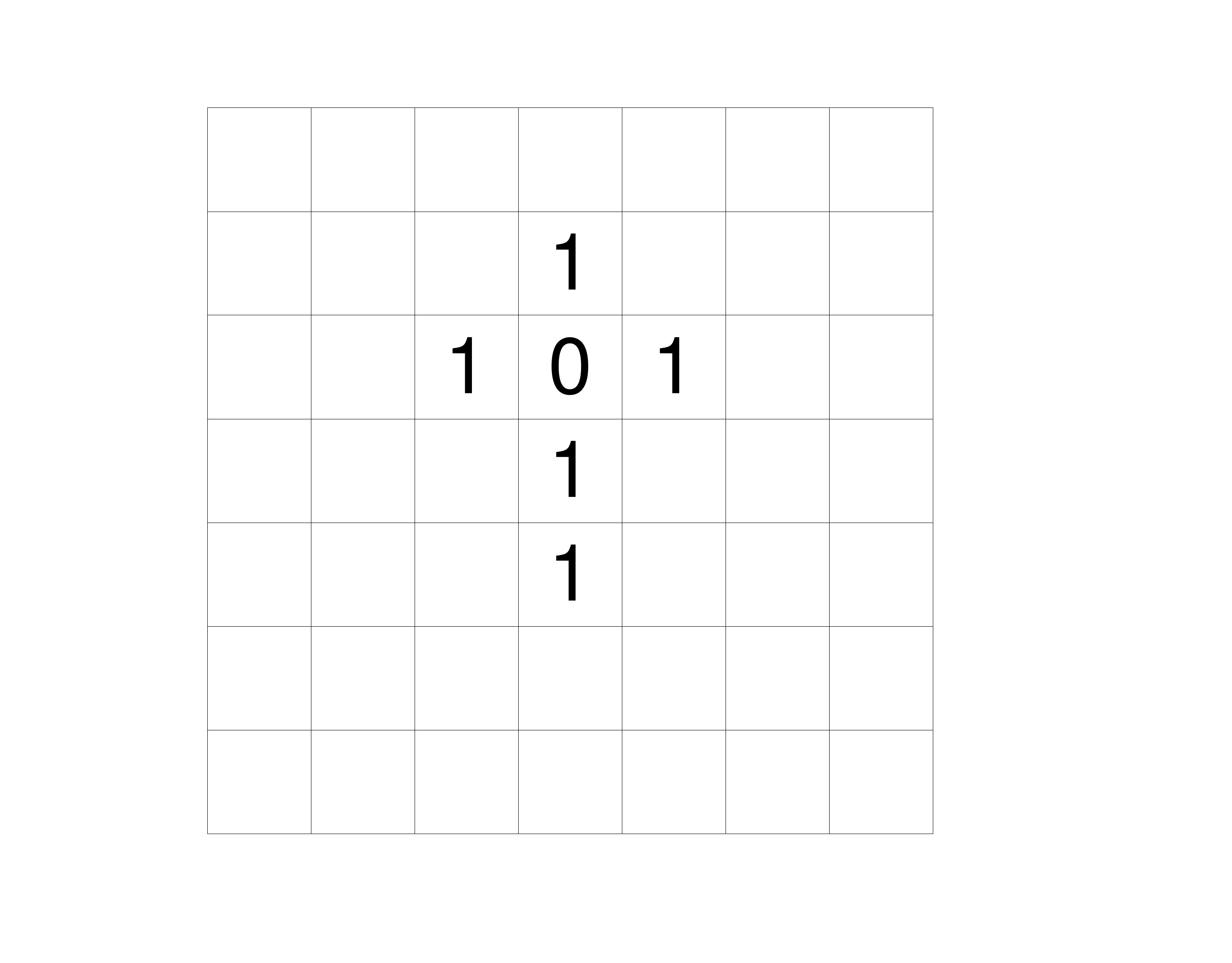} 
\includegraphics[width=0.32\textwidth]{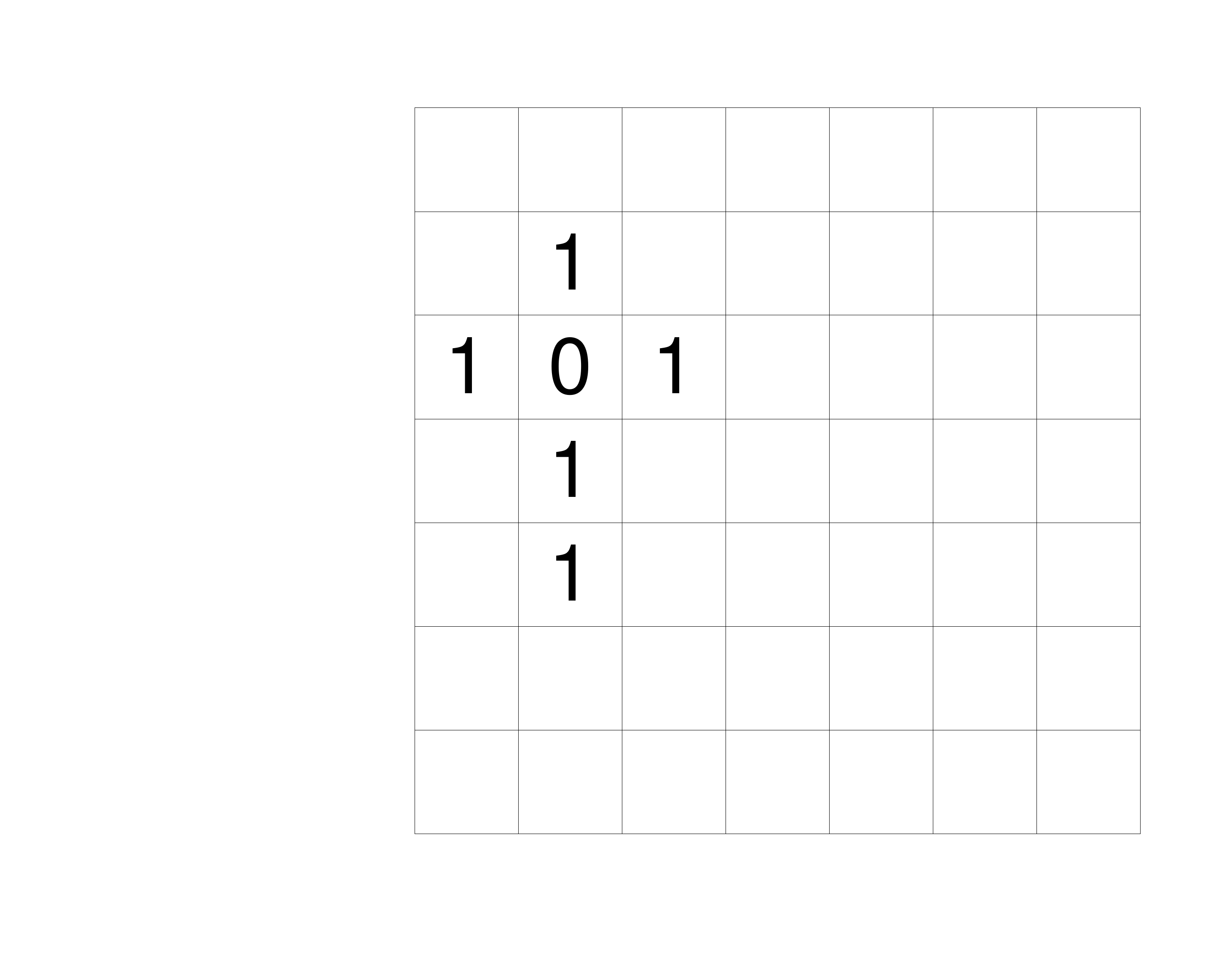} \hspace{1cm}
\includegraphics[width=0.32\textwidth]{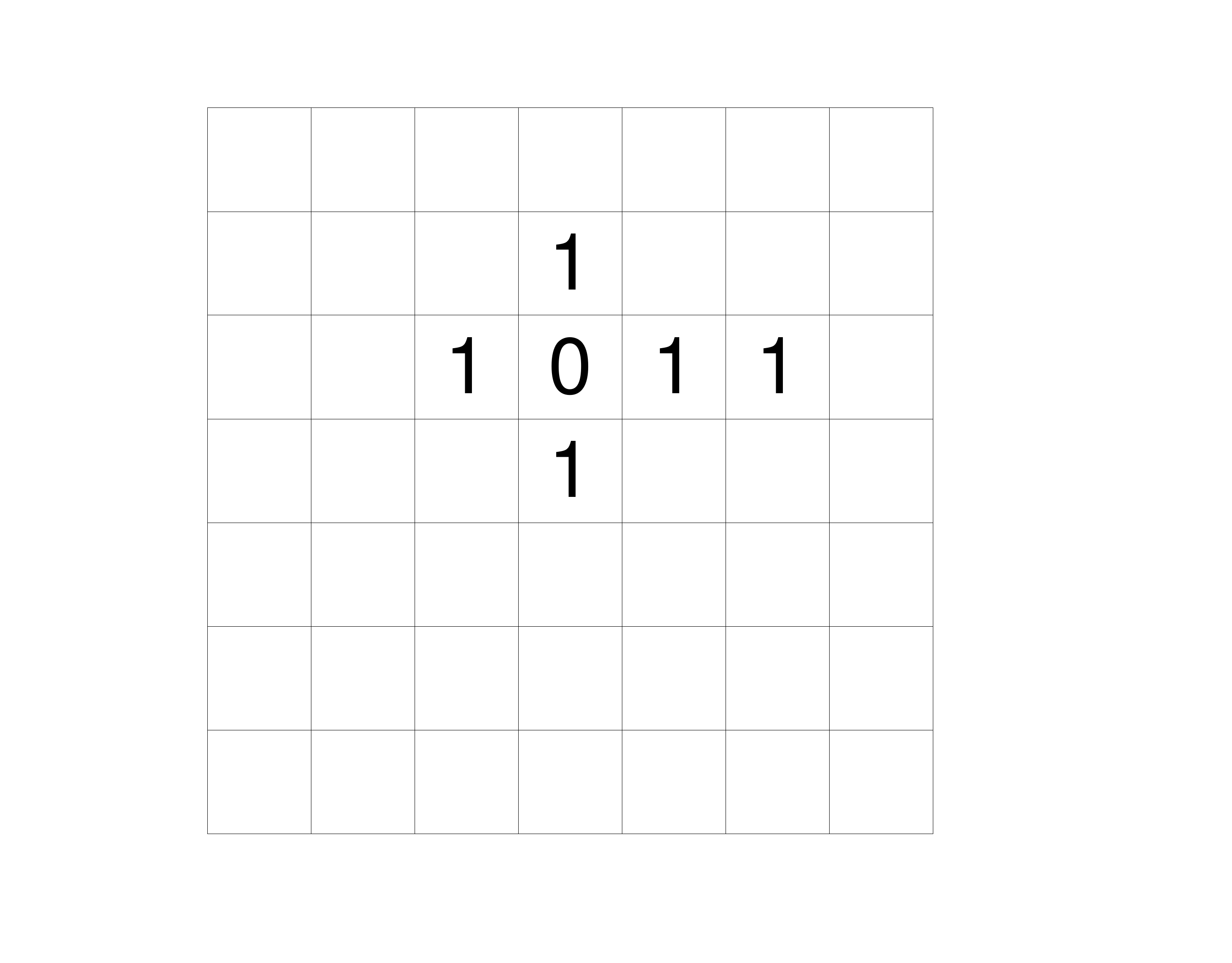} 
}
\caption{The prior probability of the configuration on the left is the same
as for its shifted copy  (middle) and its rotated copy (right) provided that
these transformations preserve   the partition into $L_+$ and  $L_-$.
For instance,  if $L_+$ is the set with $x_1>0$, then it is allowed to shift along the $x_1=0$ axis.  In 
a lattice with $d\geq 3$, there are also rotations that preserve the axis.
\label{Leaf}}
\end{figure}

Recalling our motivation for defining a prior different from Solomonoff's, 
we note that $P_t$ 
indeed respects some of the symmetries of physical laws. 
Consider, for some time $t$,  
the probability
$P_t$ of the pattern in Fig.~\ref{Leaf},  left, consisting  of symbols $1$  and $0$.
The empty squares indicate cells whose value
is unspecified. Fig.~\ref{Leaf}, middle, and right, show  shifted and rotated copies of 
the same pattern, respectively.  If the shift and the  rotation
are chosen such that they leave $L_\pm$ invariant, then $P_t$ is obviously  the same for
these copies.

To discuss item 2 in the above list of desired modifications, we assume that the generation of some 
$c$ 
requires only a short program on a Turing machine but one needs to initiale a large region $R_p$
to generate it on a physically universal CA. One reason could be that it involves a long and fragile computation process
which only outputs the correct result if a large environment is correctly initialized. Then 
$P_t(c)$ would be small for all $t$.

In the spirit of Solmonoff's prior, we would like  to 
ensure that every $c\in A^R$ (for an arbitrary finite  region $R$)  gets non-zero probability for some
$t\in \N_0$. Note that the maximally mixed state on  $L_+$ can be interpreted as 
a mixture  over ``random programs'', and it is not clear  whether programs on $L_+$ 
are sufficient for preparing any desired configuration (also on $L_-$).  
It  could be that this defines  an even stronger kind of physical  universality.
This problem will also be left open.

To define a physical analog of Kolmogorov complexity we first discuss why  the following
straightforward definition is  inappropriate for our purposes:
For any $c\in A^R$ one  could define the  complexity of $c$ as the size of the  region
$R_p$ for which there  is a state $c_p\in A^{R_p}$ such that
\[
\exists t\in \N_0 \quad \alpha_t(c_p,e)|_R =c  \quad \forall e\in A^R\,.
\]
Then the complexity of $c$  is at  least $|R|$ because $\alpha_t$ is a bijection. 
We want to define complexity of a state in such a way that simple patterns like $0^R$
have low complexity.  Unfortunately, we are not able  to show that
this will be the case for the complexity measure below, but there is at least no  obvious argument
why it cannot be (as opposed to the above measure). 
The decisive assumption
 that we make is that $R_p$ must be contained in $L_+$. This is in agreement with the fact that
our ``random programs'' 
that define $P_t$ are contained in $L_+$ while $L_-$ only contains free memory space. 

Even though we must leave it open, whether {\it every} configuration can be prepared
by programs in $L_+$ (see also the remarks above regarding the physical prior), we now define 
the ``program size complexity'', but we phrase it more general and define
the complexity of processes other  that state  preparation:
\begin{Definition}[physical complexity]${}$\\ \label{Complex}
Let $R$  be some region and
\[
M: A^R \rightarrow A^R
\] 
be an arbitrary map. The physical complexity of $M$ 
is defined by the minimum
\[
C(M):=\min \{ |R_{M}| \log_2 |A| +K(t)\}\,,
\]
where the minimum is taken over all $t\in \N$ and all regions $R_M$ and initializations $c_i \in S^{R_M}$ 
for which
\[
\alpha_t(c_M,c_i)|_{R_M} = M(c_i)\,.
\]
The physical complexity  $C(c)$  of a configuration $c$ is defined by 
the  complexity of the map $M$ with $M(c_i)=c$ for all $c_i\in A^R$.
\end{Definition}
The additional term $K(t)$ will later be needed to ensure that 
our complexity measure satisfies Kraft's inequality.
A more intuitive justification may be 
that the time parameter must be provided as {\it external} information. Since there is probably no finite initialization that prepares a state and keeps it forever, we must been told when the desired state is present or the desired transformation is performed.
The following relation between physical complexity and the physical prior is almost obvious:
\begin{Lemma}[lower bound on physical complexity]${}$\\
\begin{equation}\label{LeviAn}
C(c)\geq \min_{t\in \N_0} \{ -\log_2 P_t(c) +K(t)\}\,.
\end{equation}
\end{Lemma}

\noindent
Proof: By definition of the physical prior, 
\begin{equation}\label{PrC}
P_t(c) \geq |A|^{-|R_{p}|}\,, 
\end{equation}
for all $c_{p}$ that prepare $c$ after the time $t$.
By definition of physical complexity,
\[
C(c) =\min_{t\in \N_0} \min_{c_{p}} \{ |R_{p}| \log_2 |A| + K(t)\}\,,
\]
where the minimum is taken over all $c_{p}$ that prepare $c$ after the time $t$.
Using~(\ref{PrC}) we obtain
\begin{equation}\label{Cc}
C(c) \geq \min_{t\in \N_0} \{- \log_2 P_t(c) +K(t)\}\,.
\end{equation}
$\Box$

Rather than having inequality (\ref{LeviAn}) only one may wish to show a tighter link between
the physical prior and physical complexity -- in analogy to the tight connection
between Solomonoff's prior and Kolmogorov complexity \cite{Levin}:
\begin{Theorem}[Coding Theorem of Levin]${}$\\
\[
-\log  m(s) = K(s) +O(1)\,,
\]
where $O(1)$ means that the error can be bounded by a constant that depends on the Turing machine, but 
does not depend on $s$.
\end{Theorem}
Hence, $m(s) \approx 2^{-K(s)}$ up to a multiplicative term that is bounded by some constant. 
For this reason, tighter connections between the physical prior and physical complexity are desirable.

The following theorem describes a mathematical property of physical complexity
that it shares with Kolmogorov complexity:

\begin{Theorem}[Kraft's inequality]${}$\\ \label{Kraft}
Let $U$ be a set of mutually exclusive configurations of arbitrary  size.
Then, physical complexity satisfies 
\[
\sum_{c\in U} 2^{-C(c)}\leq 1\,.
\]
\end{Theorem}

\noindent
Proof:
Let $t_c$ be the time that minimizes the right hand side of (\ref{Cc}), hence
\[
C(c) \geq -\log_2 P_{t_c}(c) +K(t_c)\,.
\]
We conclude
\begin{eqnarray*}
\sum_{c\in U} 2^{-C(c)} &\leq & \sum_{c\in U} P_{t_c}(c) 2^{-K(t_c)}\\
                       &\leq & \sum_{t\in \N_0} \sum_{c\in U} P_t(c) 2^{-K(t)}\\
                       &\leq & \sum_{t\in \N_0} 2^{-K(t)} \leq 1\,,  
\end{eqnarray*}
where the second last inequality holds  because the configurations are mutually  exclusive and
the
 last step uses the usual Kraft inequality for Kolmogorov complexity.
$\Box$

\vspace{0.3cm}
The fact that Kolmogorov complexity satisfies Kraft's inequality (which was not the case in Kolmogorov's version since he did not use prefix codes) made it possible to renormalize
it to a probability distribution on strings, yielding Solomonoff's prior.

Although a better understanding of our notion of physical complexity has to be left to the future,
it is, by construction, clear that it takes into account whether running a process  requires
to adjust a large part of the environment -- even though the process may be simple from the point of
view of algorithmic information. 
Such a strong disagreement between  Kolmogorov complexity and physical complexity occurs e.g. if $R_p$ is large but $c_p$ mainly consists of zeros, or some other
algorithmically simple pattern.
If a physical process requires, for instance, cooling  a large region (e.g. setting many cells to zero) around the system this could formally appear as large
physical complexity.

\section{Physical universality in the quantum world}

\label{Sec:quantum}

\subsection{Informal description of some differences to the classical case}

The main question that  arises when  we translate the notion of universal state
preparation into the quantum world is  whether the configuration of the environment 
is supposed to be
a {\it basis} state. In other words, we ask 
whether the preparation of general quantum superposition should be reducible to the
preparation of basis states in the environment.

On the one hand, it seems to be  artificial to select a certain subset of states
as being more fundamental than others.
On the other  hand, the following model 
suggests that basis states should  be sufficient:
we could think of the basis  states  as states in the  register of a classical processor that controls a quantum preparation machine.  Then the register is the region that  we act on by changing its {\it classical}
state only.

\subsection{Defining the problem}

To formally define quantum CAs, we assume that every site $x\in L$ contains a quantum system 
with Hilbert space $\cH:=\C^a$, where $a:=|A|$ and the basis vectors $|j\rangle$ are  labelled by
symbols $j\in A$. The Hilbert space of a region $R$ is then given by  the tensor product  
of copies of $\cH$, but to avoid problems with infinite tensor products 
we follow \cite{GrossWerner}
and use an operator algebraic framework \cite{BR1,BR2}:
Let every site $x$ be described by a copy  of the same matrix  algebra $\cA_x$   of $a\times a$ matrices. The self-adjoint part of $\cA_x$ is interpreted   
as the observables corresponding to cite $x$.
For every finite set $\Lambda\in L$, let $\cA_\Lambda$ be the tensor product
\[
\cA_\Lambda :=\bigotimes_{x\in \Lambda} \cA_x\,.
\]
For $\Lambda \subset \Lambda'$, $\cA_{\Lambda}$ is considered as 
subalgebra of $\cA_{\Lambda'}$ in a canonical way by adding the tensor product of an appropriate number 
of $a\times a$ identity matrices.
For every infinite set $\Lambda$, we define $\cA_\Lambda$ as  the $C^*$-completion 
over the union of algebras of finite regions $\cA_{\Lambda_f}$. 
This defines the $C^*$-algebra $\cA_L$ 
which contains all local algebras\footnote{the ``quasi-local'' algebra \cite{BR1}}.

The set $S(\cA_L)$ of states
is the set of positive linear functionals $\phi: \cA_L \rightarrow  \C$
with $\phi({\bf 1 })=1$. The  state space 
$S(\cA_L)$ is a convex set whose extreme points are called pure states, this 
definition generalizes density operators of rank one to the infinite system.
A pure state $\phi$ is said to be a basis state on a region  $R$ if 
it is given by
\[
\phi (a)={\tt tr}(\rho a)  \quad\forall a\in \cA_R\,,
\] 
where $\rho$ is a {\it diagonal} matrix with diagonal $(0,\dots,0,1,0,\dots,0)$.
A pure state is said to be a (global) basis state if its restriction to every finite region
is a basis state.

It is convenient to describe  the dynamics in the Heisenberg picture, it is 
then given by a group $(\alpha_t)$ of $C^*$-automorphisms
of $\cA_L$ satisfying the following locality condition:
\[
\phi (\cA_x) \in  \cA_R\,,
\] 
for every region $R$  that contains the Moore  neighborhood of $x$ with radius one.
The dynamics transfers the state $\phi$ into $\phi \circ \alpha_t$.
For any observable $a\in \cA_R$ for which $\alpha_t (a)\in \cA_{R'}$ for some region $R'$,  
the value 
$(\phi \circ \alpha_t) (a)$ is already determined by the restriction of $\phi$ to
$R'$. Therefore, 
$(\rho \circ \alpha_t) (a)$ is also a well-defined expression if $\rho$ is a state on $\cA_{R'}$.

The following  notion of physical universality 
can be seen as a  quantum analog of Definition~\ref{Cstate} to the quantum world. 
As opposed to the set of classical configurations  of 
a finite region, the set of pure states is (uncountably) infinite. On  the other hand, the set of
basis states of a region $R_p$  is finite and the ste  of all basis  states of the whole lattice still is countable,
we cannot prepare all states on $R$ exactly but at most up to any desired accuracy:

\begin{Definition}[conditional quantum state preparation]${}$\\ \label{qUn}
A  quantum CA is said to allow for conditional state preparation
if for every 
pair of states  $(\rho_i,\rho_f) \in S(\cA_R)\times S(\cA_R)$ of a region  $R$ and every $\epsilon>0$
there  is a basis state $\gamma  \in S(\cA_{L\setminus R})$ of the complement and a time $t$ such that
\[
|(\gamma \otimes \rho_i)\circ \alpha_t (a) -\rho_f(a)| \leq \epsilon \|a\| \quad  \forall a\in \cA_R\,,
\]
where $\|.\|$ denotes the operator norm.
\end{Definition}

It is important to note that the program state $\gamma$ is a basis state, i.e.,
the program is {\it classical} software.
As opposed to the classical case, this  notion of universality is not satisfied by
the ``trivial'' CA that  only shifts the state.
Instead,  it includes problems like how to prepare sophisticated 
multi-particle entanglement using a given interaction via preparing the environment
to {\it basis} states.
We thus formulate the following  open problem:

\begin{Question}[physically universal  quantum CA]${}$\\ \label{ExQ}
Is there a quantum CA that is physically universal in the sense of
Definition~\ref{qUn}? 
\end{Question}

We will not translate Definitions~\ref{Ustate} and \ref{bijections}
to the quantum setting since even our ``weak'' form of universality is not obvious to exist for quantum CAs.

\subsection{Physically universal Hamiltonians}

\label{Subsec:Hamil}

To account for the fact that time evolutions are actually continuous,
we may want to switch from CAs to Hamiltonians.
In the literature there exists a large number of translation invariant finite range
Hamiltonians on lattices that are universal for quantum computing, e.g., \cite{Ergodic,ErgodicQutrits,Aha_spinchain,Vollbrecht}, but physical universality  
has not been considered. A characteristic feature of many constructions for computational universal Hamiltonians is the separation between a ``program region'' and a ``data region'' where 
the former  controls the operations performed on the latter. Physical universality 
would  imply that we are also able to operate on  the program region, which could  require
an infinite hierarchy of program regions. 
To formally define physical universality,
we can straightforwardly adapt Definition~\ref{qUn} 
by replacing the group $(\alpha_t)_{t\in \Z}$  with the continuous version 
$(\alpha_t)_{t\in  \R}$. 
To properly state what it means that a dynamics of an infinite lattice is given
by a finite range translation invariant Hamiltonian
we consider an operator $h \in \cA_R$ for some region $R$ and 
define for every vector $x\in \Z^d$, the shifted copy of $h$ by $\tau_x(h)$. 
Then it is known that the differential equation
\begin{equation}\label{Schroedinger}
\frac{d}{dt} \alpha_t(a) = i\left[\sum_{x\in \Z^d} \tau_x(h) ,a\right]
\end{equation}
defines uniquely  a group of $C^*$-automorphisms \cite{BR2}.
Definition~\ref{qUn} and, correspondingly, Question~\ref{ExQ} then straightforwardly 
translate to the group $\alpha_t$ defined by (\ref{Schroedinger}). 

The considerations on the thermodynamic costs change more significantly because 
we replace the maximum entropy state by the state of minimum free energy, i.e.,
the Gibbs state (for defining thermal equilibrium states for infinite lattices see \cite{BR2}), which ensures that we are getting  closer to real physics. We may then even allow for lattices having an 
infinite dimensional algebra at each site.
We also want to translate the physical prior and the physical complexity
in Subsection~\ref{Subsec:Kolmo}.
Now, the notion of hot and cold parts is taken more literally than above since
the definition of Hamiltonians allows us to defined thermal states for temperatures other than
$T=0$ and $T=\infty$.
Thermal equilibrium states on infinite quantum lattice systems can be defined via limits of Gibbs states for finite regions \cite{BR2} (we do not care about  the potential non-uniqueness of limit points here).
We restrict these states of the infinite lattice to 
$L_+$ and $L_-$, respectively and ``glue'' them together
 to define our initial state:

\begin{Definition}[initial state of the universe]${}$\\
Let $\phi_T: \cA_L\rightarrow \C$ be Gibbs  states for temperature $T$ 
on the entire lattice.
For some $T_2>T_1>0$, let
$\phi_+$ be  the  restriction of  $\phi_{T_2}$ to  $\cA_{L_+}$ and $\phi_-$ the restriction of
$\phi_{T_1}$ to $\cA_{L_-}$.  Then we  define the ``initial state of the universe'' by
\[
\phi:=\phi_+ \otimes \phi_-\,.
\]
\end{Definition}

\begin{Definition}[physical prior for Hamiltonian systems]${}$\\ \label{HamP}
For every $t$ we define the mixed state
\[
\phi_t:=\phi \circ \alpha_t\,.
\]
Let $|\psi\rangle$ be the state vector of some pure state on $\cA_R$.
Then
\[
\phi_t(|\psi\rangle \langle \psi|)
\]
is the probability for obtaining the state $|\psi\rangle \langle \psi|$ after the time $t$
when measuring a non-degenerate self-adjoint operator that contains $|\psi\rangle$ as one of its eigenvectors.
\end{Definition}

In  the spirit of Solomonoff's prior, we would like to give higher prior to states that
are simple in an intuitive sense than to complex ones. 
For instance, we  would consider the basis state
$|0\rangle \langle 0|^R$ (i.e., all cells in the region $R$ 
are in the state $|0\rangle \langle 0|$) 
as simple.
It is possible that a small program makes the Hamiltonian dynamics
generating free memory space via using the temperature gradient. This is at least not forbidden by
any obvious thermodynamic laws.
Thermodynamics also allows for processes that use the existing temperature gradient
to either lower the temperature of some region in $L_-$ (refrigerator driven by a heat engine, see also \cite{JWZGB}) or increase the temperature of $L_+$ even further. 
The size of the program required to make $\alpha_t$ implementing such a  process would then  be
the physical complexity of the process.
This is only meant to be one of many examples how physically universal CAs define the complexity
of physical processes, no matter whether they are computation processes or not.

An interesting   modification of the above would be given by 
replacing the lattice with a  field-theoretic model, where nets of subalgebras $\cA_\Lambda$
are assigned to regions in $\R^d$ \cite{Ha92} and define physical universality for a field theory. 
As opposed to the discrete model, this would allow for 
the definition of an even ``more physical'' prior that is  invariant under the full Lorentz group.

\section{Conclusions}

The main contribution of this paper is to introduce and motivate the concept 
of physically universal CAs and Hamiltonians. Their non-existence 
would probably have interesting consequences for the  limits of
controlling microscopic systems. 
But also  their existence poses questions that are equally fundamental, because such CAs
are nice models for studying the thermodynamic cost of 
computation and control. 

We also use physically universal CAs to define the complexity
of states and a corresponding prior probability that is considered as a physically motivated
analog of Solomonoff's prior. An interesting feature of this prior is that
it is invariant under some physical symmetries. Moreover, it
tries to capture the amount of adjustments that is needed in the environment
to run a preparation process,  which includes also the cost of removing 
disturbing heat and the  cost of keeping it away from the system during the
implementation of the
process.

\vspace{0.3cm}

The author would like to  thank Bastian Steudel and David  Balduzzi for helpful  comments on an
earlier draft and Aram Harrow and Armen Allahverdyan for interesting discussions.
This work has partially been supported by the VW-project ``Quantum Thermodynamics:
energy and  information flow at nanoscale''.

%\bibliographystyle{unsrt}
%\bibliography{../../../literatur/literatur}

\end{document}